\documentclass[a4paper,11pt]{article}
\pdfoutput=1
\usepackage{pos}
\usepackage[firstpage]{draftwatermark}
\SetWatermarkFontSize{12pt}
\SetWatermarkColor{black}
\SetWatermarkAngle{0}
\SetWatermarkHorCenter{19.4cm}
\SetWatermarkVerCenter{5.8cm}
\SetWatermarkText{MS-TP-22-56}
\usepackage{macros}
\usepackage{dsfont}
\usepackage{subfigure}
\usepackage{booktabs}
\usepackage{sidecap}
\title{Sigma terms of the baryon octet in $N_\mathrm{f} = 2+1$ QCD\\ with Wilson quarks}
\ShortTitle{Sigma terms of the octet baryons}

\author*[a]{Pia Leonie Jones Petrak}
\author[b]{Gunnar Bali}
\author[b]{Sara Collins}
\author[a]{Jochen Heitger}
\author[b]{Daniel Jenkins}
\author[b]{Simon Weish\"aupl}

\affiliation[a]{
	Institut für Theoretische Physik, Westfälische Wilhelms-Universität M\"unster,\\
	Wilhelm-Klemm-Straße 9, 48149  M\"unster, Germany}

\affiliation[b]{Institut für Theoretische Physik, Universität Regensburg, 93040 Regensburg, Germany}
\emailAdd{p\_petr04@uni-muenster.de}
\emailAdd{gunnar.bali@ur.de,sara.collins@ur.de}
\emailAdd{heitger@uni-muenster.de}
\emailAdd{Daniel.Jenkins@ur.de}
\emailAdd{Simon.Weishaeupl@ur.de}

\abstract{A lot of progress has been made in the direct determination of nucleon sigma terms. 
	Using similar methods, we consider the sigma terms of the other octet baryons as well. These are determined on CLS gauge field ensembles employing the Lüscher-Weisz gluon action and the Sheikholeslami-Wohlert fermion action with $N_\mathrm{f} = 2 + 1$. The ensembles have pion masses ranging from ${410}\,\mathrm{MeV}$ down to the physical value and lattice spacings covering a range between 
	${0.098}\,\mathrm{fm}$ and ${0.039}\,\mathrm{fm}$. We present some preliminary results for the pion and strange sigma terms and compare to indirect determinations. To do so, we discuss  multi-state fits to tackle the well-known problem of excited state contamination comparing the ratio and summation methods also including priors.}

\FullConference{%
The 39th International Symposium on Lattice Field Theory,\\
8th-13th August, 2022,\\
Rheinische Friedrich-Wilhelms-Universität Bonn, Bonn, Germany
}

\begin{document}

\maketitle

\section{Introduction}
The sigma terms encode the quark contributions to the mass of a baryon. They are defined as the matrix element of a scalar current $J$ times the quark mass such that
\begin{align}
\sigma_{qB} = m_q \langle B| J |B \rangle\,,
\label{eq:sigma_term}
\end{align}
where $J = \bar{q} \,\mathds{1} \, q$ and the quark flavour $q\in \{u,d,s\}$. In order to compare with
phenomenological determinations, the pion-baryon sigma terms are usually
constructed, $\sigma_{\pi B} = \sigma_{uB} + \sigma_{dB}$. In the matrix element, $B$ refers to the ground state of a baryon $B$.
Of particular interest are the nucleon sigma terms ($B=N$) which appear in the expressions for WIMP-nucleon scattering cross-sections and are relevant for comparing model predictions
to the exclusion bounds obtained from direct detection dark matter experiments (such as the XENON1T experiment).

In the analysis, we make use of and adjust methods established for the
nucleon (see ref.~\cite{Ottnad:2020qbw} for a review).  We determine the sigma terms of
the baryon octet, i.e. of the lambda $\Lambda$, sigma $\Sigma$ and cascade $\Xi$
baryons. This enables us to investigate SU(3) flavour symmetry
breaking and to determine the SU(3) low energy constants (LECs), that
are currently not well known. There have been relatively few
determinations so far, see, for example, refs.~\cite{Shanahan:2012wh} and \cite{Durr:2011mp}, with most
previous studies focusing on the nucleon sigma terms~(see ref.~\cite{FlavourLatticeAveragingGroupFLAG:2021npn} for
a review).


In addition, discrepancies between results for the pion-nucleon sigma term from Lattice QCD and phenomenology are still to be resolved (see \cite{FlavourLatticeAveragingGroupFLAG:2021npn}, and e.g., \cite{Alexandrou:2019brg,Borsanyi:2020bpd,Hoferichter:2016ocj}). In a recent paper, results more consistent with phenomenology were obtained by explicitly including $N\pi$ and $N\pi\pi$ excited states in the analysis \cite{Gupta:2021ahb,Gupta:2022aba}. By considering baryons other than the nucleon, we aim to understand
excited state contributions and other issues in more detail so as to
help solve this puzzle.

\section{Excited state analysis}
In order to be able to extract the ground-state matrix element, we
need to take care of the excited state contamination. Two possible
approaches are the ratio method and the summation method (reviewed in \cite{Green:2018vxw,Ottnad:2020qbw}) that are both based on spectral decompositions. We consider the two- and three-point functions of a baryon (from the octet) at rest in the initial and final state.
The spectral decomposition of the two-point function reads
\begin{align}
C_\mathrm{2pt}(\tf)=\sum_{\vec{x}}\left\langle \mathcal{O}_\mathrm{snk}(\vec{x},\tf) \bar{\mathcal{O}}_\mathrm{src}(\vec{0},0) \right\rangle
= \sum_n |Z_n|^2 e^{-E_n \tf}\, ,
\end{align}
where $Z_n\propto\langle\Omega|\mathcal{O}_\mathrm{snk}|n\rangle$ is the overlap of the interpolator $\mathcal{O}_\mathrm{snk}$ onto the state $n$ (and $\Omega$ the vacuum state) and $\tf$ the source-sink separation. Summation over spin and colour indices and projection onto
positive parity are implied. These indices become apparent when writing down the operators explicitly. The interpolators for the four octet baryons are set to
\begin{align}
	\mathcal{O}_\mathrm{snk}^{\alpha,\mathrm{N}} &= \epsilon^{abc} u_{a}^\alpha \left( u_{b}^\beta (C\gamma_5)^{\beta\gamma}d_c^\gamma\right) \quad \text{and} \quad
	\mathcal{O}_\mathrm{snk}^{\alpha,\Lambda} = \epsilon^{abc} s_{a}^\alpha \left( u_{b}^\beta (C\gamma_5)^{\beta\gamma}d_c^\gamma\right),\nonumber \\
	\mathcal{O}_\mathrm{snk}^{\alpha,\Sigma} &= \epsilon^{abc} u_{a}^\alpha \left( u_{b}^\beta (C\gamma_5)^{\beta\gamma}s_c^\gamma\right) \quad \,\text{and} \quad
	\mathcal{O}_\mathrm{snk}^{\alpha,\Xi} = \epsilon^{abc} s_{a}^\alpha \left( s_{b}^\beta (C\gamma_5)^{\beta\gamma}u_c^\gamma\right).
\end{align}
$a,b,c$ are colour indices, $\alpha,\beta,\gamma$ are spin indices and  $\mathcal{O}_\mathrm{src}^\alpha = \mathcal{O}_\mathrm{snk}^\alpha$ and $\bar{\mathcal{O}}_\mathrm{src} = \mathcal{O}_\mathrm{src}^\dagger \gamma_4 $. $C$ stands for the charge conjugation operator. Note that for the $\Lambda$ we use a naive interpolator that also has overlap with the (heavier) $\Sigma^0$. Turning to the three-point function, its spectral decomposition reads
\begin{align}
C_\mathrm{3pt}(\tf,t) &=\sum_{\vec{x},\vec{y}}\left\langle \mathcal{O}_\mathrm{snk}(\vec{x},\tf) J(\vec{y},t) \bar{\mathcal{O}}_\mathrm{src}(\vec{0},0) \right\rangle
- \sum_{\vec{x},\vec{y}} \left\langle J(\vec{y},t)\right\rangle\left\langle \mathcal{O}_\mathrm{snk}(\vec{x},\tf)\nonumber \bar{\mathcal{O}}_\mathrm{src}(\vec{0},0) \right\rangle\\
&=\sum_{n,n'} Z_{n'} Z_n^* \langle n'|J|n\rangle
e^{-E_nt} e^{-E_{n'}(\tf-t)},
\label{eq:C3pt}
\end{align}
where $t$ is the insertion time of the scalar current, $J = \bar{q} \, \mathds{1} \, q$. 
As $J$ has the same quantum numbers as the vacuum, the vacuum expectation value needs to be subtracted, see the second term in the first line of eq.~(\ref{eq:C3pt}). 
Note that depending on the type of baryon and current, different Wick contractions (e.g. different currents) contribute that result in connected and disconnected quark-line diagrams. 

\label{sect:excited_state_analysis}
\\{\bf{Ratio method:}}\\
Taking the ratio of the two spectral decompositions and truncating
after the first excited state contribution leads to
\begin{align}
R(\tf,t) = \frac{C_\mathrm{3pt}(\tf,t)}{C_\mathrm{2pt}(\tf)} = g_S^q + c_{01} \mathrm{e}^{-\Delta \, \cdot \,t} + c_{10} \mathrm{e}^{-\Delta \, \cdot \, (\tf-t)} + c_{11} \mathrm{e}^{-\Delta \, \cdot \, \tf} + ...
\label{eq:ratio_method}
\end{align}
where $g_S^q =\langle B|\bar{q} \, \mathds{1} \, q| B\rangle $ is the ground-state matrix element of interest. $\Delta = E_1  - E_0$ is the energy gap between the ground and first excited state. The coefficients $c_{01}$, $c_{10}$, $c_{11}$ are made up of matrix elements for different transitions, $B_0 \rightarrow B_1$, $B_1 \rightarrow B_0$ ,  and
$B_1 \rightarrow B_1$, respectively. Here, $B_0$ refers to the ground state of the baryon while $B_1$ denotes the first excited state (single- or multi-particle state). As the baryon is at rest, $c_{01} = c_{10} \equiv c_{0\leftrightarrow1}$. Note that we labelled $c_{0\leftrightarrow1}$ as $c_1$ in \cite{Petrak:2021aqf}.
\\{\bf{Summation method:}}\\
Summing over the ratio (\ref{eq:ratio_method}) for a range of insertion times leads to
\begin{align}
\sum_{t/a=c}^{{\small t_\mathrm{f}/a-c}} R(t_\mathrm{f},t) =g_S^q (t_\mathrm{f}/a -2c +1) + \frac{2c_{0\leftrightarrow1}}{1-\mathrm{e}^{a\Delta}}
\left(\mathrm{e}^{\Delta(c-t_\mathrm{f})} - \mathrm{e}^{\Delta(a-c)} \right)
+ c_{11} ({\small t_\mathrm{f}/a} -2c +1)  \mathrm{e}^{-\Delta  \, \cdot \, t_\mathrm{f}} +{\small ...}
\label{eq:summation_method}
\end{align}
where $a$ refers to the lattice spacing and $c>0$ to preserve reflection positivity (we set $c=2$). An advantage of this method is that, in principle, the slope, {\small $\frac{\mathrm{d}\quad \,}{\mathrm{d}(t_\mathrm{f}/a)}\sum_{t/a=c}^{t_\mathrm{f}/a-c} R(t_\mathrm{f},t) = g_S^q + \mathrm{O}(t_\mathrm{f}/a \,\cdot \, \mathrm{e}^{-\Delta  \, \cdot \,  t_\mathrm{f}})\,,$}
approaches the asymptotic value faster compared to the ratio method. However, a large number of source-sink separations is required and due to the the numerical setup we use (discussed in the next
section) we can only employ this approach for the disconnected three-point functions.
\section{Numerical Setup}
\label{sect:numerical_setup}
\begin{SCfigure}
\centering
\includegraphics[width=0.54\textwidth]{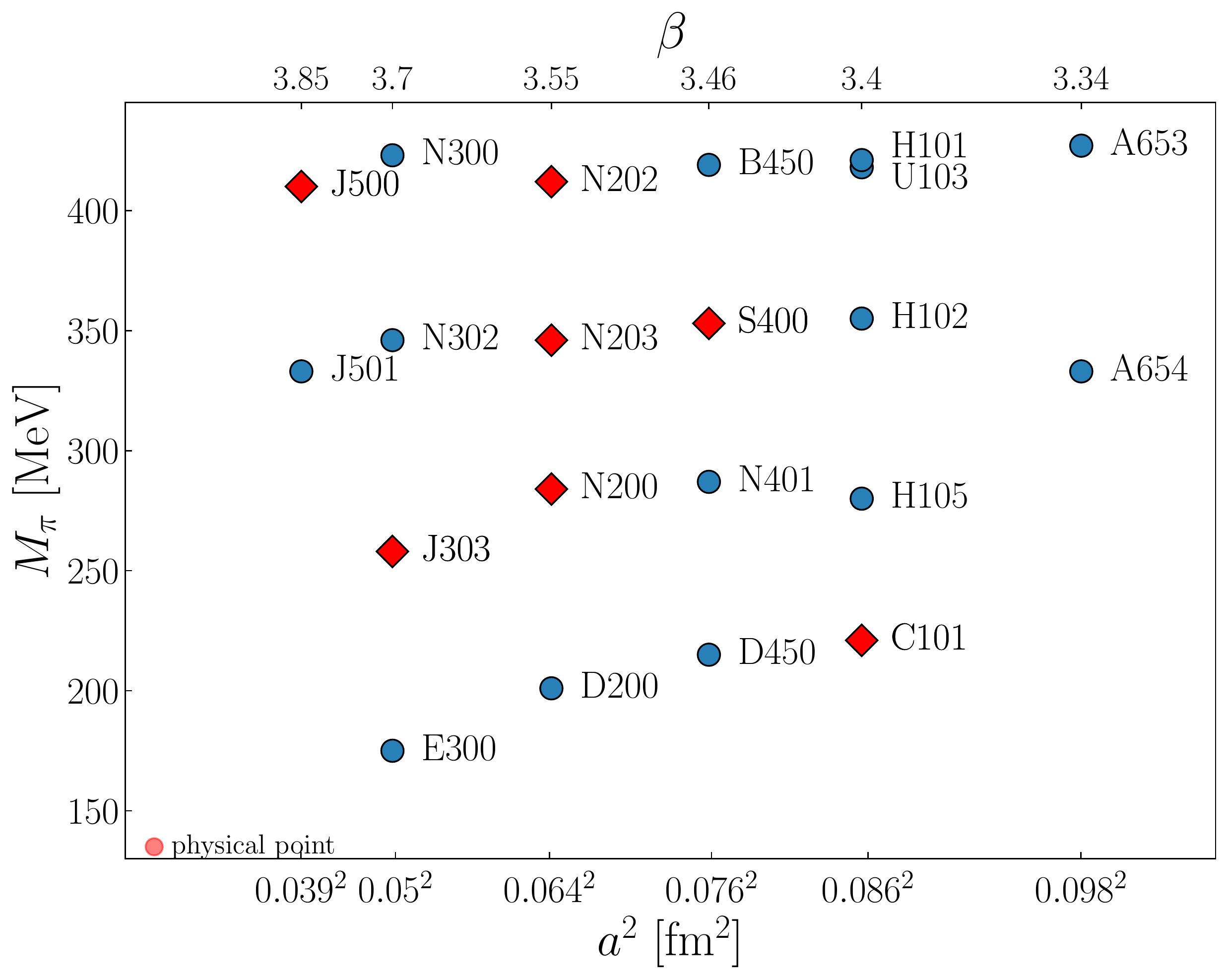}
\caption{Overview of a subset of the CLS ensembles which lie on a trajectory
which starts at a symmetric point~($m_l=m_s$, $M_\pi\sim 410$~MeV)
and extends to the physical point, along which the flavour average of
the quark masses is held constant.  So far the ensembles highlighted
in red (diamonds) have been analysed.}
\label{fig:ensembles}
\end{SCfigure}

We utilise the CLS gauge field ensembles \cite{Bruno:2014jqa}, which are generated
employing the Lüscher-Weisz gluon action and the
Sheikholeslami-Wohlert fermion action with $N_\mathrm{f} = 2 + 1$ ($m_l=m_u=m_d\leq m_s$). 
The pion-baryon and strange sigma terms are determined on the seven
ensembles highlighted in red~(diamonds) in fig.~\ref{fig:ensembles}.  This includes five
lattice spacings in the range from
$a=0.08528(49)$~fm down to $a=0.03875(22)$~fm. The pion masses vary from 411 MeV down to
222 MeV, see also~\cite{RQCD:2022xux} 
\footnote{In \cite{Petrak:2021aqf} we investigated the quark mass dependence by performing a preliminary chiral extrapolation at $a\approx0.064\,\mathrm{fm}$ (including preliminary SU(3) LECs from \cite{Bali:2022qja}).}.


To compute the connected three-point correlation functions on the
ensembles with $m_l = m_s$ and $M_\pi=410$~MeV, we used the standard
sequential source method  \cite{Maiani:1987by}. On the other ensembles we employed the
stochastic method described in \cite{Bali:2019svt,Bali:2017mft} (see also \cite{Yang:2015zja,Alexandrou:2013xon,Bali:2013gxx,Evans:2010tg}). This
approach enables us to obtain measurements for all baryons of interest
at multiple source and insertion positions, simultaneously. Four
different source-sink separations, corresponding to $\tf \approx [0.71 \,\mathrm{fm}, 0.9 \,\mathrm{fm}, 1.03 \, \mathrm{fm}, 1.22 \, \mathrm{fm}]$, are employed. Two measurements are
performed for each $\tf$ on every configuration except for the $m_l = m_s$
ensembles (that we used the sequential source method for) where
ten measurements are typically performed (one, two, three and four
measurements for  $\tf \approx [0.71 \,\mathrm{fm}, 0.9 \,\mathrm{fm}, 1.03 \, \mathrm{fm}, 1.22 \, \mathrm{fm}]$, respectively).

The disconnected three-point functions are constructed by correlating a quark loop with a baryon two-point function. The loop is estimated stochastically on every timeslice (within the bulk) leading to
additional noise on top of the Monte-Carlo gauge sampling. In order to reduce the noise, the truncated solver method~\cite{Bali:2009hu}, the hopping parameter expansion technique~\cite{Thron:1997iy} and time partitioning~\cite{Bernardson:1993he} are utilised. 
Between twenty and thirty measurements of the two-point functions (at
different temporal source positions) are performed on each
configuration.  A reasonable signal for the disconnected three-point
function is obtained for a source-sink separation of up to $\tf \approx 1.22\,\mathrm{fm}$.  
For the analysis of the statistical errors we employ the $\Gamma$-method \cite{Wolff:2003sm} (that is based on autocorrelation functions) using the \textsf{pyerrors python} package \cite{Joswig:2022qfe}.

When constructing the sigma terms for Wilson-type fermions, one needs
to take the mixing between different quark flavours into account, which arises
due to the singlet and non-singlet combinations of the scalar current
renormalising differently as a result of chiral symmetry breaking (see , e.g., \cite{Petrak:2021aqf} for the renormalisation pattern). For the ratio of renormalisation factors, $r_\mathrm{m}$, we use $r_\mathrm{m}(\beta=3.4)=2.335(31)$, $r_\mathrm{m}(\beta=3.46)=1.869(19)$, $r_\mathrm{m}(\beta=3.55)=1.523(14)$, $r_\mathrm{m}(\beta=3.7)=1.267(16)$ and $r_\mathrm{m}(\beta=3.85)= 1.149(18)(27)[33]$, determined  non-perturbatively in \cite{Heitger:2021bmg}.
\section{Fitting Analysis}
\begin{figure}
	\centering
	\includegraphics[width=0.49\linewidth]{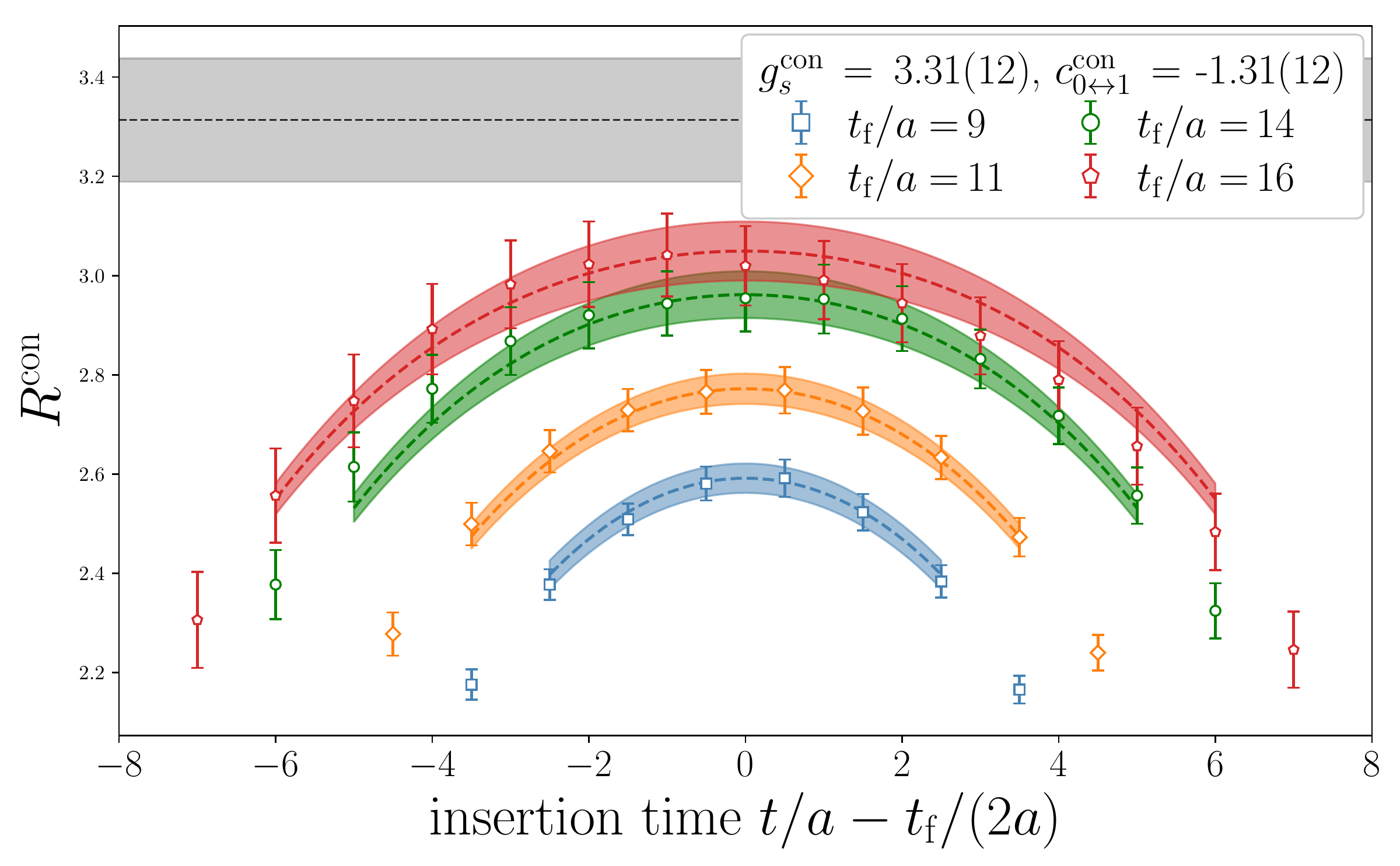}
	\includegraphics[width=0.49\linewidth]{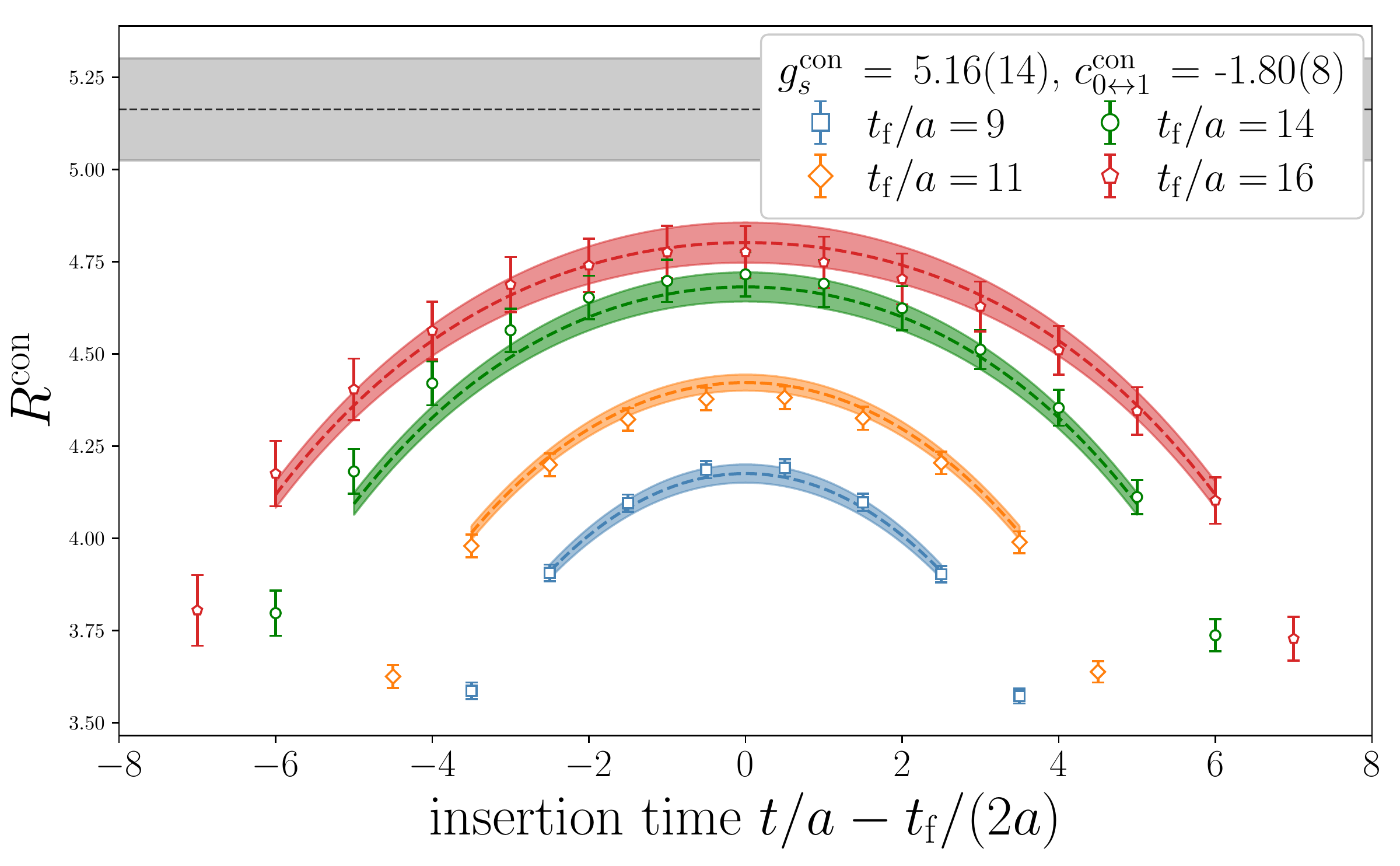}
	
	\includegraphics[width=0.49\linewidth]{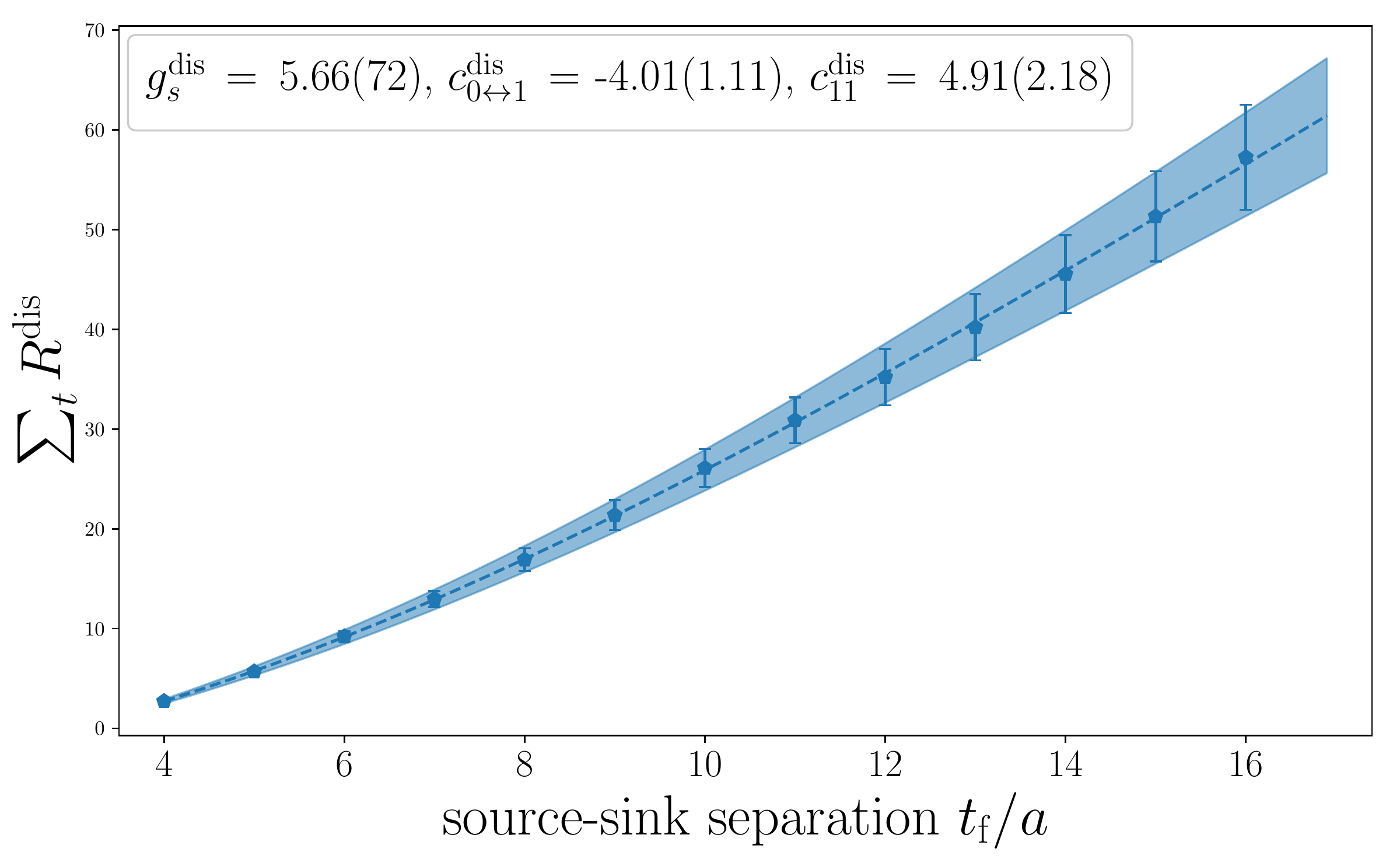}
	\includegraphics[width=0.49\linewidth]{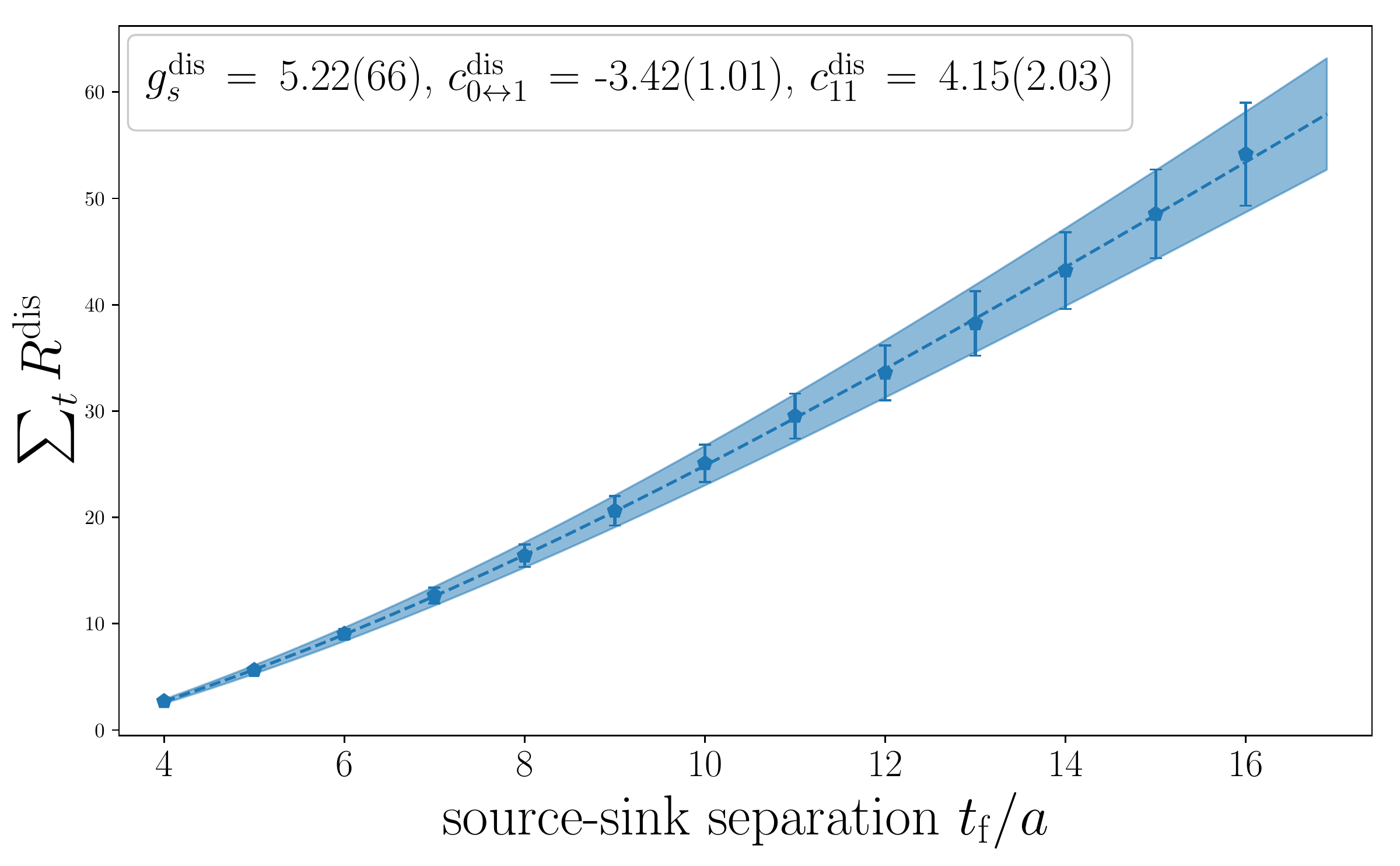}
	\caption{ 
			The connected and (summed) disconnected ratios contributing to the
			unrenormalised scalar charges of the $\Xi$ baryon on ensemble S400~($M_\pi\approx 350$~MeV)
			for the $\bar{u}u$ current (left) and the $\bar{s}s$ current (right): Simultaneous
			fits to the connected and summed disconnected ratios are indicated by
			the coloured shaded regions. The resulting connected ground state scalar matrix
			elements are displayed as horizontal grey bands.}
	\label{fig:summation_sim_fit}
\end{figure}

In order to extract the ground-state matrix elements of interest while
taking the excited state contamination into account we perform
multi-state fits, applying the ratio method, eq.~(\ref{eq:ratio_method}) and, for
comparison, for the disconnected contribution also the summation
method, eq.~(\ref{eq:summation_method}) (as noted in sect.~\ref{sect:excited_state_analysis}, there are not enough
source-sink separations available to apply this method to the
connected contributions). For each baryon, we fit the connected and
disconnected ratios for all currents simultaneously, with the energy
gap to the first excited state $\Delta$ as the common fit parameter. As an
example, the ratios (and fits) relevant for determining the sigma
terms of the $\Xi$ baryon on the S400 ensemble~($M_\pi\approx 350$~MeV)
are displayed in fig.~\ref{fig:summation_sim_fit} (see fig. 2 in~\cite{Petrak:2021aqf} for an example of the ratio method being used for both the connected and disconnected contributions). While we were able to resolve the first
excited state term, with the coefficient $c_{0\leftrightarrow1}$, with both methods, the
$B_1\to B_1$ transition term~(see sect.~\ref{sect:excited_state_analysis}) was not resolved when
employing the ratio method and we set $c_{11} = 0$ in the analysis
when using this method. The $\chi^2/\chi_\mathrm{expected}^2$ were mostly below 1.5 (and always below two), where $\chi_\mathrm{expected}^2$ provides an estimate of the effective degrees of
freedom expected taking into account autocorrelations, see \cite{Bruno:2022mfy}.
\begin{figure}
	\centering
	\includegraphics[width=1.0\linewidth]{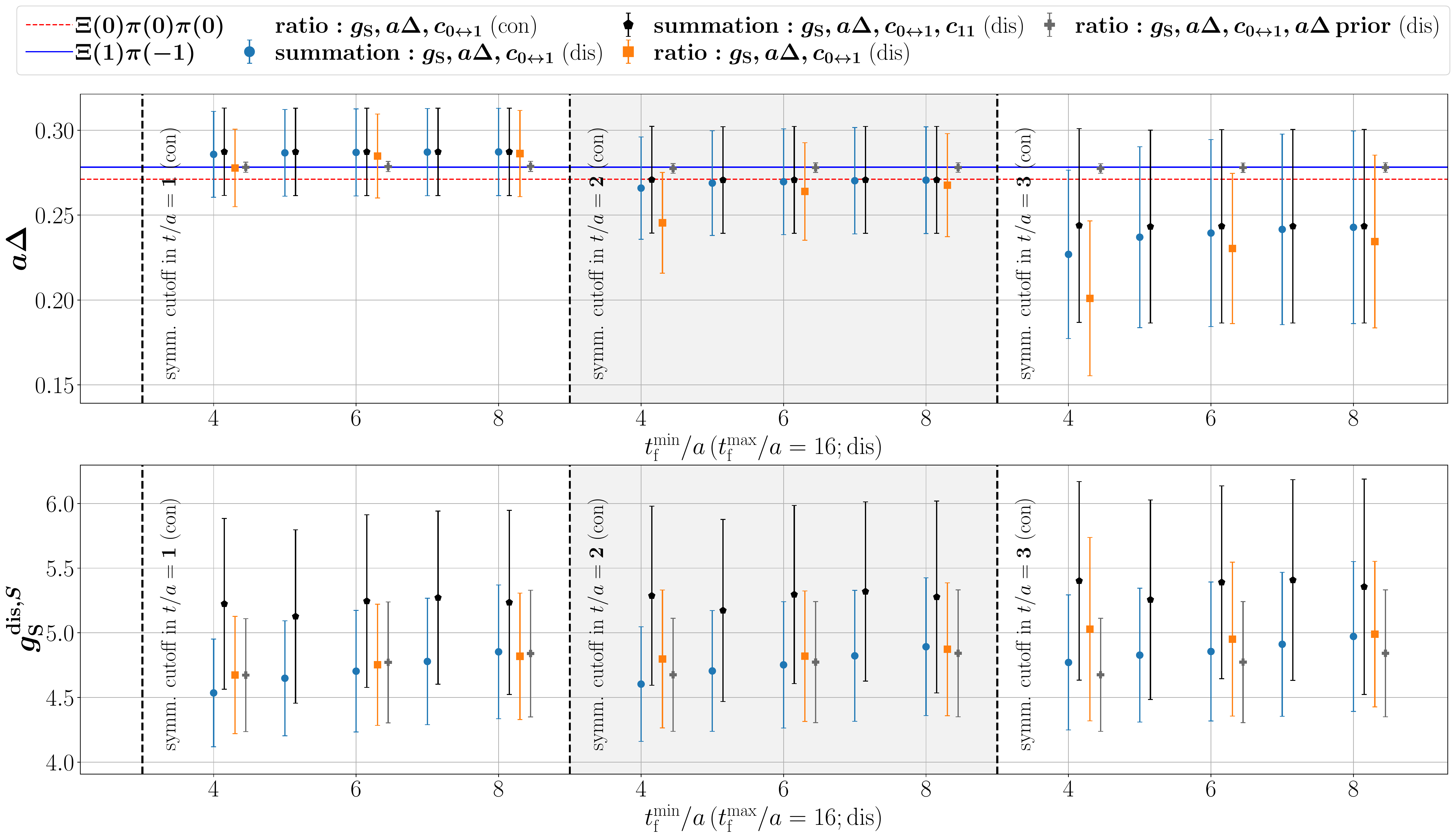}
	\caption{
			Effect of the fit form and range variation on the energy
			gap $\Delta= E_1 - E_0$ (top) and the disconnected
			scalar charge $g_S^\mathrm{dis}$ for $J=\bar{s}s$ (bottom) for the $\Xi$
			baryon on the ensemble S400~($M_\pi\approx 350$~MeV).  The blue and
			red-dashed horizontal lines indicate an energy gap corresponding to a
			$\Xi(1)\pi(-1)$ and $\Xi(0)\pi(0)\pi(0)$ excited state, respectively. The numbers in brackets stand
			for the momentum as a multiple of the lowest lattice momentum $2\pi/L$. Each of the
			three panels displays results of simultaneous fits to the connnected
			and disconnected contributions where the fit range is fixed for the
			former and varied for the latter, while across the different panels,
			the fit range for the connected ratio is varied. 
			The insertion times included are always symmetric w.r.t. the source and sink.
			This leads to a minimal $\tf$ for the fits to the disconnected ratios~(for which all $\tf$
			are, in principle, available). For the connected contributions only
			the ratio method (with $c_{11}=0$) is employed, while for the
			disconnected results both the ratio and the summation
			method are shown. Results for the ratio method, using a prior (Gaussian with $\sigma=1\%$) for the
			energy gap corresponding to a $\Xi(1)\pi(-1)$
			(non-interacting) excited state, are also displayed. The
			first (black) pentagon corresponds to the fit in fig.~\ref{fig:summation_sim_fit}.}
	\label{fig:variation}
\end{figure}


To investigate whether the excited state contributions are
sufficiently controlled when extracting the ground state matrix
elements, we vary the fit range for both the connected and
disconnected contributions and for the latter also the fit method.  In
addition, we consider the impact on the fit of a narrow-width prior
for $\Delta$. The fit range is varied by removing insertion times
symmetrically for each source-sink separation. The results for the
example of S400 are displayed in fig.~\ref{fig:variation}. We find that the energy gap
is mostly determined by the connected data: in the first row of fig.~\ref{fig:variation}
the error on $\Delta$ increases as more connected data points are
excluded from the fit. We also observe that the energy gap $\Delta= E_1 - E_0$  is
compatible with that for a ${\Xi(1)\pi(-1)}$
or $\Xi(0)\pi(0)\pi(0)$ first excited
state so that a narrow-width prior corresponding to a ${\Xi(1)\pi(-1)}$ excited state has almost no effect on the ground-state matrix elements (second row of fig.~\ref{fig:variation}).
This is also true for the other members of the baryon octet (when comparing to the energy of a ${B(1)\pi(-1)}$
or $B(0)\pi(0)\pi(0)$ excited state).  However, as the pion mass decreases, with the exception of
the nucleon, we find $E_1$ to be larger than the energies of these levels. 
This may be related to the fact that the spectrum (for the nucleon) is denser at smaller pion masses \cite{Green:2018vxw}
making it more difficult to separate the first excited state contributions from others. 
However, we can enforce
the first excitation to have the energy of a non-interacting $B(1)\pi(-1)$
state 
by using a narrow-width prior and  the sigma terms are usually only mildly affected, see fig.~\ref{fig:fit_form_comparison}.
\\
\section{Preliminary results and consistency with indirect determinations}
\begin{figure}
	\centering
	\includegraphics[width=1.0\linewidth]{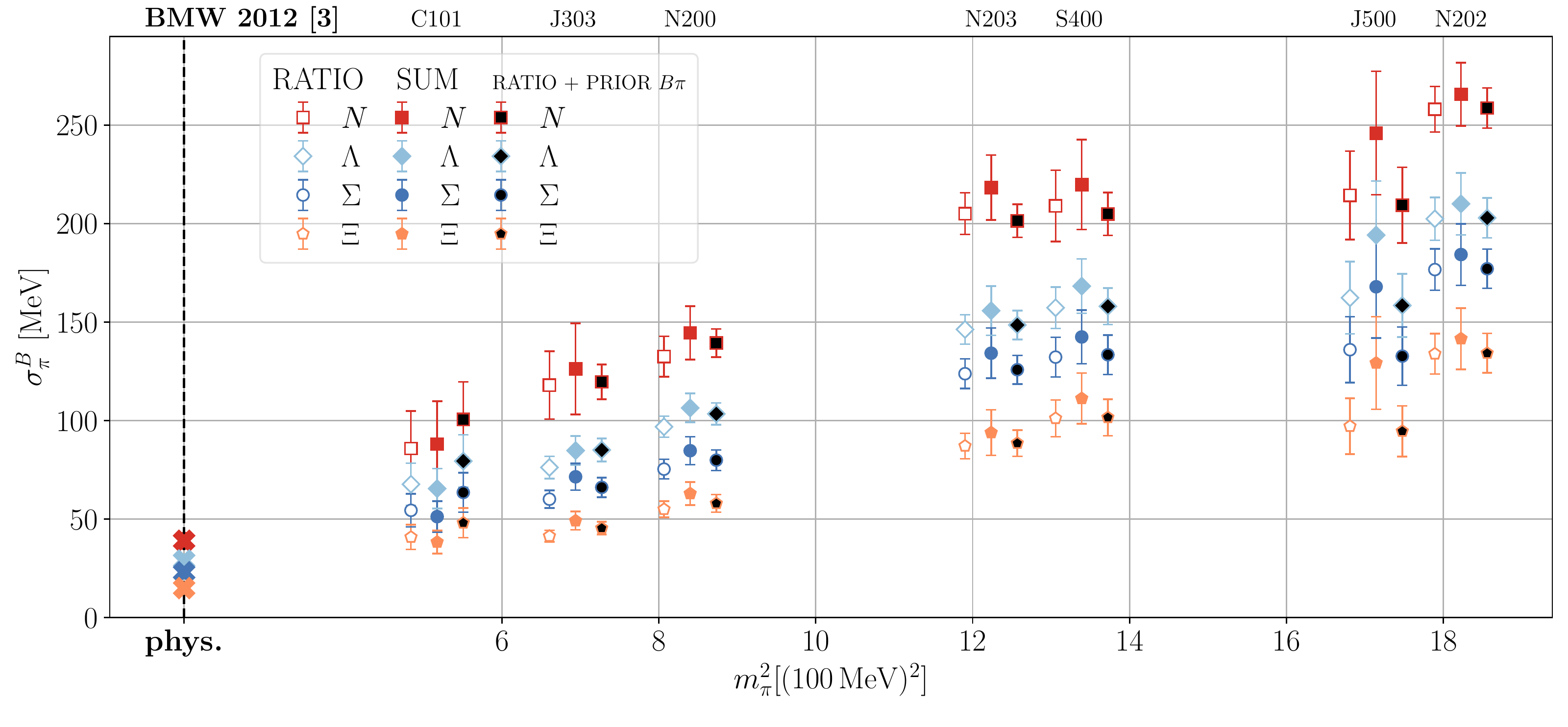}
	\includegraphics[width=1.0\linewidth]{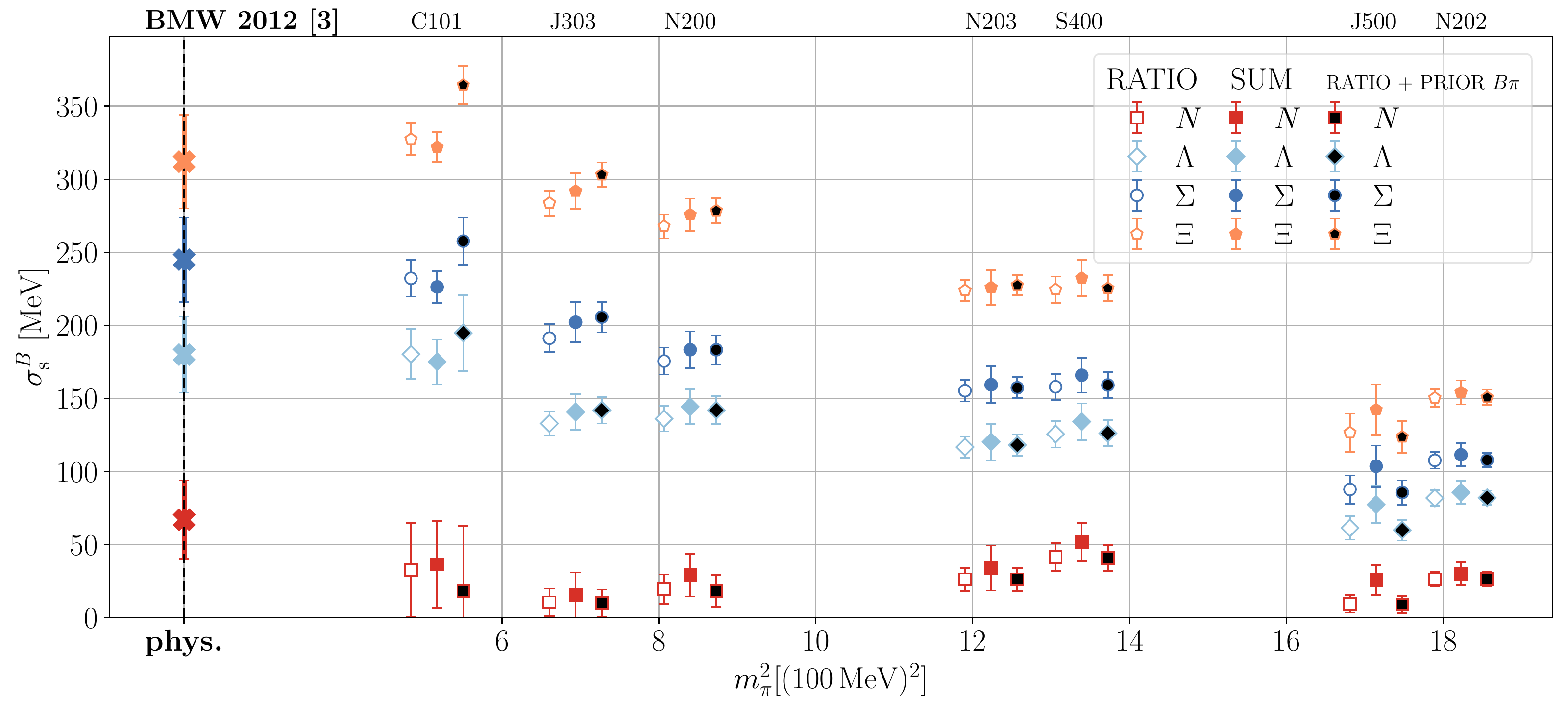}
	\caption{
	Preliminary results for the pion-baryon and strange sigma
	terms as a function of the pion mass squared obtained utilising the ratio
	method (with $c_{11}=0$) for both contributions (with and without a
	prior for the energy gap) and utilising the ratio method for the
	connected contribution and the summation method (with $c_{11}\neq0$) for the disconnected
	contribution. The prior width is set to $1\%$ centered around an
	energy gap compatible with a ${B(1)\pi(-1)}$ excited state. We also compare our results to those of BMW 2012~\cite{Durr:2011mp}.}
	\label{fig:fit_form_comparison}
\end{figure}

\begin{figure}
	\centering
	\includegraphics[width=0.49\linewidth]{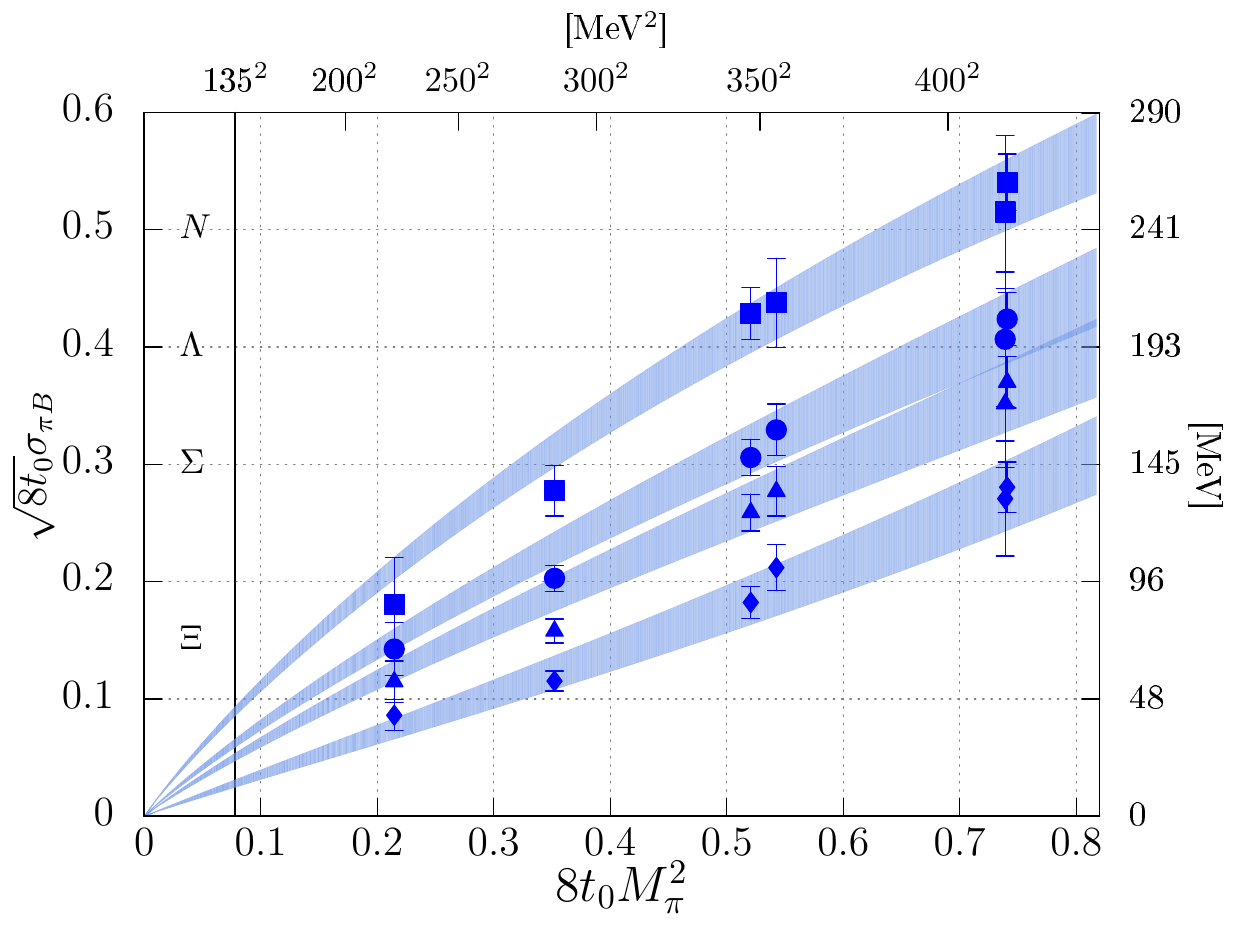}
	\includegraphics[width=0.49\linewidth]{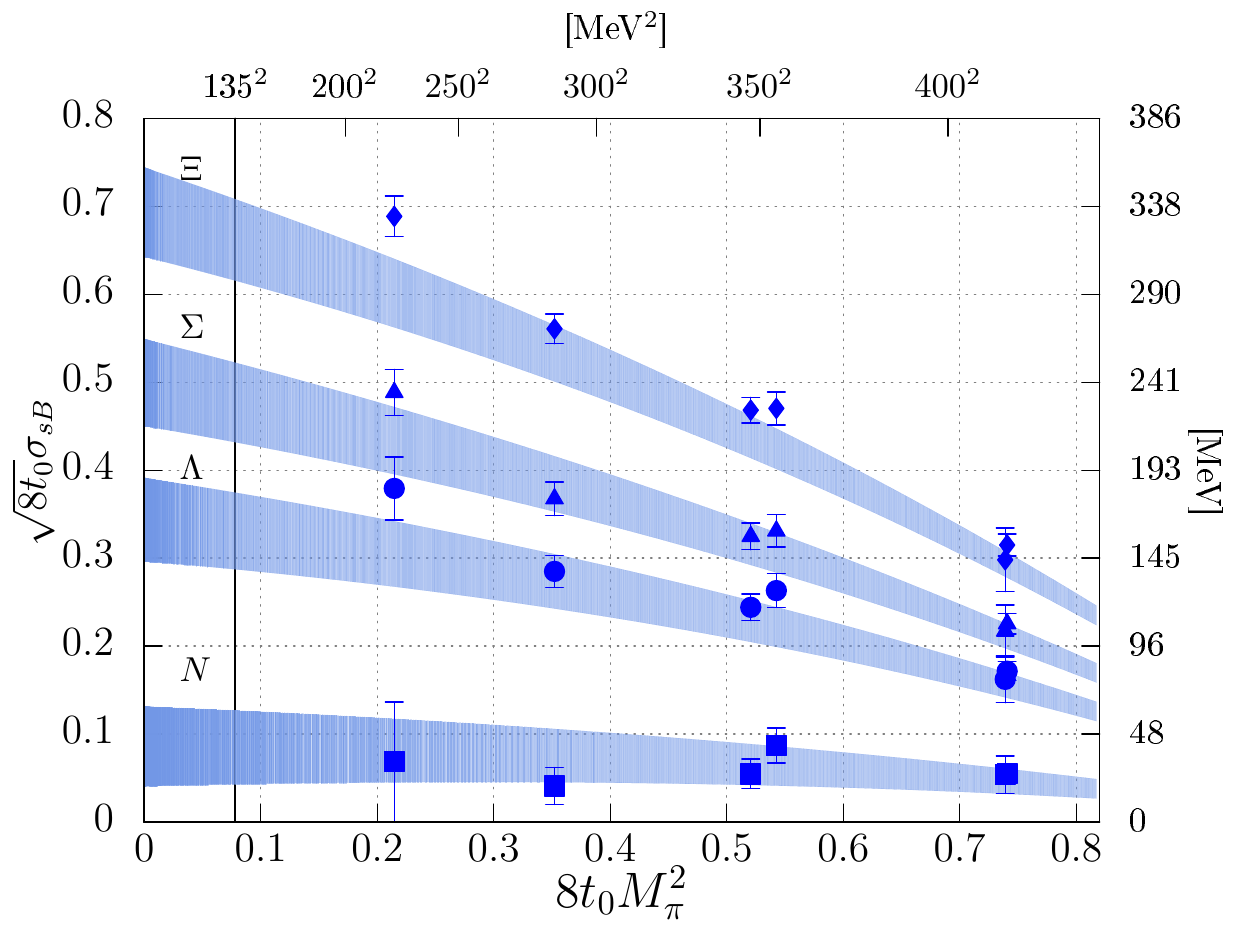}
	\caption{
	Results for the pion-baryon and strange sigma terms as a
	function of the pion mass squared obtained using the ratio method
	(blue squares). The shaded regions show the quark mass dependence of
	$\sigma_\pi^B$ and $\sigma_s^B$ determined via the Feynman-Hellmann theorem utilising a fit
	to the octet baryon masses evaluated on 47 CLS
	ensembles~\cite{RQCD:2022xux}. Finite-volume effects are accounted
	for and NNLO baryon chiral perturbation theory is used to describe the quark mass
	dependence.}
	\label{fig:indirect}
\end{figure}

The ground-state matrix elements of interest are extracted from the
fits presented in the previous section and combined with the
corresponding quark masses and renormalisation factors~(see \cite{Petrak:2021aqf}) to
form the pion-baryon and strange sigma terms for all four octet
baryons.  Our preliminary results are displayed as a function of the
pion mass squared in fig.~\ref{fig:fit_form_comparison}. We see that the different fitting methods
employed lead to compatible results for the sigma terms.  In addition,
we compare our results from the ratio method (extracted at finite
lattice spacing) in fig.~\ref{fig:indirect} to the quark mass dependence in the
continuum limit obtained from an indirect determination via the
Feyman-Hellmann theorem and a fit to the octet baryon masses, see~\cite{RQCD:2022xux}.  Almost all our preliminary results (from our direct
determination) lie within the error bands of the indirect
determination suggesting that
, after having taken the continuum and infinite volume limits, the results
will remain consistent.

\section{Conclusion and outlook}
Our aim is to determine the strange and pion sigma terms for the
baryon octet. In these proceedings we reported on the progress made in
controlling excited state contaminations. We find that the ratio and
summation methods give compatible results for the sigma
terms. Although these methods do not always result in a first excited
state compatible with a ${B(1)\pi(-1)}$ or $B(0)\pi(0)\pi(0)$ state (for octet
baryons other than the nucleon), when a narrow-width prior is used for
the energy gap, consistent results for the sigma terms are, nonetheless,
obtained. The next steps will include analysing additional ensembles
so as to allow for an extrapolation to the physical point and an
investigation of cut-off and finite-volume effects.

\acknowledgments
This work is supported by the Deutsche Forschungsgemeinschaft (DFG) through the Research Training Group ``GRK 2149: Strong and Weak Interactions -- from Hadrons to Dark Matter'' (P. L. J. P. and J. H.). 
G. B., S. C., D. J. and S. W. were supported by the European Union’s Horizon 2020 research and innovation programme under the Marie Skłodowska-Curie grant agreement no. 813942 (ITN EuroPLEx) and grant agreement no. 824093 (STRONG-2020). S.C. and S.W. received support through the German Research Foundation
(DFG) grant CO 758/1-1.\\
\indent We gratefully acknowledge computing time granted by the
John von Neumann Institute for Computing (NIC), provided on the Booster
partition of the supercomputer JURECA~\cite{jureca} at
\href{http://www.fz-juelich.de/ias/jsc/}{J\"ulich Supercomputing Centre (JSC)}.
Additional simulations were carried out at the QPACE~3
Xeon Phi cluster of SFB/TRR~55.
The authors also gratefully acknowledge the Helmholtz Data Federation (HDF) for funding this work by providing services and computing time on the HDF Cloud cluster at the Jülich Supercomputing Centre (JSC)~\cite{hdfcloud}.
\bibliographystyle{JHEP}
\setlength{\bibsep}{0pt minus 0.8ex}
\bibliography{bib}

\providecommand{\href}[2]{#2}\begingroup\raggedright\begin{thebibliography}{10}

\bibitem{Ottnad:2020qbw}
K.~Ottnad, \emph{{Excited states in nucleon structure calculations}},
  \href{https://doi.org/10.1140/epja/s10050-021-00355-5}{\emph{Eur. Phys. J. A}
  {\bfseries 57} (2021) 50} [\href{https://arxiv.org/abs/2011.12471}{{\ttfamily
  2011.12471}}].

\bibitem{Shanahan:2012wh}
P.~E. Shanahan et~al., \emph{{Sigma terms from an SU(3) chiral extrapolation}},
  \href{https://doi.org/10.1103/PhysRevD.87.074503}{\emph{Phys. Rev. D}
  {\bfseries 87} (2013) 074503}
  [\href{https://arxiv.org/abs/1205.5365}{{\ttfamily 1205.5365}}].

\bibitem{Durr:2011mp}
S.~{D\"{u}rr} et~al., \emph{{Sigma term and strangeness content of octet
  baryons}}, \href{https://doi.org/10.1103/PhysRevD.85.014509}{\emph{Phys. Rev.
  D} {\bfseries 85} (2012) 014509}
  [\href{https://arxiv.org/abs/1109.4265}{{\ttfamily 1109.4265}}], [Erratum:
  Phys.Rev.D 93, 039905 (2016)].

\bibitem{FlavourLatticeAveragingGroupFLAG:2021npn}
{\scshape Flavour Lattice Averaging Group (FLAG)} collaboration, Y.~Aoki
  et~al., \emph{{FLAG Review 2021}},
  \href{https://doi.org/10.1140/epjc/s10052-022-10536-1}{\emph{Eur. Phys. J. C}
  {\bfseries 82} (2022) 869}
  [\href{https://arxiv.org/abs/2111.09849}{{\ttfamily 2111.09849}}].

\bibitem{Alexandrou:2019brg}
C.~Alexandrou et~al., \emph{{Nucleon axial, tensor, and scalar charges and
  $\sigma$-terms in lattice QCD}},
  \href{https://doi.org/10.1103/PhysRevD.102.054517}{\emph{Phys. Rev. D}
  {\bfseries 102} (2020) 054517}
  [\href{https://arxiv.org/abs/1909.00485}{{\ttfamily 1909.00485}}].

\bibitem{Borsanyi:2020bpd}
S.~Borsanyi et~al., \emph{{Ab-initio calculation of the proton and the
  neutron's scalar couplings for new physics searches}},
  \href{https://arxiv.org/abs/2007.03319}{{\ttfamily 2007.03319}}.

\bibitem{Hoferichter:2016ocj}
M.~Hoferichter et~al., \emph{{Remarks on the pion\textendash{}nucleon
  \ensuremath{\sigma}-term}},
  \href{https://doi.org/10.1016/j.physletb.2016.06.038}{\emph{Phys. Lett. B}
  {\bfseries 760} (2016) 74}
  [\href{https://arxiv.org/abs/1602.07688}{{\ttfamily 1602.07688}}].

\bibitem{Gupta:2021ahb}
R.~Gupta et~al., \emph{{Pion\textendash{}Nucleon Sigma Term from Lattice QCD}},
  \href{https://doi.org/10.1103/PhysRevLett.127.242002}{\emph{Phys. Rev. Lett.}
  {\bfseries 127} (2021) 242002}
  [\href{https://arxiv.org/abs/2105.12095}{{\ttfamily 2105.12095}}].

\bibitem{Gupta:2022aba}
R.~Gupta et~al., \emph{{The pion-nucleon sigma term from Lattice QCD}},  in
  \emph{{10th International workshop on Chiral Dynamics}}, 3, 2022,
  \href{https://arxiv.org/abs/2203.13862}{{\ttfamily 2203.13862}}.

\bibitem{Green:2018vxw}
J.~Green, \emph{{Systematics in nucleon matrix element calculations}},
  \href{https://doi.org/10.22323/1.334.0016}{\emph{PoS} {\bfseries LATTICE2018}
  (2018) 016} [\href{https://arxiv.org/abs/1812.10574}{{\ttfamily
  1812.10574}}].

\bibitem{Petrak:2021aqf}
P.~L.~J. Petrak et~al., \emph{{Towards the determination of sigma terms for the
  baryon octet on $N_\mathrm{f} = 2 + 1$ CLS ensembles}},
  \href{https://doi.org/10.22323/1.396.0072}{\emph{PoS} {\bfseries LATTICE2021}
  (2022) 072} [\href{https://arxiv.org/abs/2112.00586}{{\ttfamily
  2112.00586}}].

\bibitem{Bruno:2014jqa}
M.~Bruno et~al., \emph{{Simulation of QCD with $N_\mathrm{f} = 2 + 1$ flavors
  of non-perturbatively improved Wilson fermions}},
  \href{https://doi.org/10.1007/JHEP02(2015)043}{\emph{JHEP} {\bfseries 02}
  (2015) 043} [\href{https://arxiv.org/abs/1411.3982}{{\ttfamily 1411.3982}}].

\bibitem{RQCD:2022xux}
{\scshape RQCD} collaboration, G.~S. Bali et~al., \emph{{Scale setting and the
  light baryon spectrum in $N_f=2+1$ QCD with Wilson fermions}},
  \href{https://arxiv.org/abs/2211.03744}{{\ttfamily 2211.03744}}.

\bibitem{Bali:2022qja}
{\scshape RQCD} collaboration, G.~S. Bali et~al., \emph{{Leading order mesonic
  and baryonic SU(3) low energy constants from $N_\mathrm{f} = 3$ lattice
  QCD}}, \href{https://doi.org/10.1103/PhysRevD.105.054516}{\emph{Phys. Rev. D}
  {\bfseries 105} (2022) 054516}
  [\href{https://arxiv.org/abs/2201.05591}{{\ttfamily 2201.05591}}].

\bibitem{Maiani:1987by}
L.~Maiani et~al., \emph{{Scalar Densities and Baryon Mass Differences in
  Lattice {QCD} With Wilson Fermions}},
  \href{https://doi.org/10.1016/0550-3213(87)90078-2}{\emph{Nucl. Phys. B}
  {\bfseries 293} (1987) 420}.

\bibitem{Bali:2019svt}
G.~S. Bali et~al., \emph{{Hyperon couplings from $N_\mathrm{f} = 2 + 1$ lattice
  QCD}}, \href{https://doi.org/10.22323/1.363.0099}{\emph{PoS} {\bfseries
  LATTICE2019} (2019) 099} [\href{https://arxiv.org/abs/1907.13454}{{\ttfamily
  1907.13454}}].

\bibitem{Bali:2017mft}
G.~S. Bali et~al., \emph{{Baryonic and mesonic 3-point functions with open spin
  indices}}, \href{https://doi.org/10.1051/epjconf/201817506014}{\emph{EPJ Web
  Conf.} {\bfseries 175} (2018) 06014}
  [\href{https://arxiv.org/abs/1711.02384}{{\ttfamily 1711.02384}}].

\bibitem{Yang:2015zja}
Y.-B. Yang et~al., \emph{{Stochastic method with low mode substitution for
  nucleon isovector matrix elements}},
  \href{https://doi.org/10.1103/PhysRevD.93.034503}{\emph{Phys. Rev. D}
  {\bfseries 93} (2016) 034503}
  [\href{https://arxiv.org/abs/1509.04616}{{\ttfamily 1509.04616}}].

\bibitem{Alexandrou:2013xon}
{\scshape ETM} collaboration, C.~Alexandrou et~al., \emph{{A Stochastic Method
  for Computing Hadronic Matrix Elements}},
  \href{https://doi.org/10.1140/epjc/s10052-013-2692-3}{\emph{Eur. Phys. J. C}
  {\bfseries 74} (2014) 2692}
  [\href{https://arxiv.org/abs/1302.2608}{{\ttfamily 1302.2608}}].

\bibitem{Bali:2013gxx}
G.~S. Bali et~al., \emph{{Nucleon structure from stochastic estimators}},
  \href{https://doi.org/10.22323/1.187.0271}{\emph{PoS} {\bfseries LATTICE2013}
  (2014) 271} [\href{https://arxiv.org/abs/1311.1718}{{\ttfamily 1311.1718}}].

\bibitem{Evans:2010tg}
R.~Evans et~al., \emph{{Improved Semileptonic Form Factor Calculations in
  Lattice QCD}}, \href{https://doi.org/10.1103/PhysRevD.82.094501}{\emph{Phys.
  Rev. D} {\bfseries 82} (2010) 094501}
  [\href{https://arxiv.org/abs/1008.3293}{{\ttfamily 1008.3293}}].

\bibitem{Bali:2009hu}
G.~S. Bali et~al., \emph{{Effective noise reduction techniques for disconnected
  loops in Lattice QCD}},
  \href{https://doi.org/10.1016/j.cpc.2010.05.008}{\emph{Comput. Phys. Commun.}
  {\bfseries 181} (2010) 1570}
  [\href{https://arxiv.org/abs/0910.3970}{{\ttfamily 0910.3970}}].

\bibitem{Thron:1997iy}
C.~Thron et~al., \emph{{Pade - Z(2) estimator of determinants}},
  \href{https://doi.org/10.1103/PhysRevD.57.1642}{\emph{Phys. Rev. D}
  {\bfseries 57} (1998) 1642}
  [\href{https://arxiv.org/abs/hep-lat/9707001}{{\ttfamily hep-lat/9707001}}].

\bibitem{Bernardson:1993he}
S.~Bernardson et~al., \emph{{Monte Carlo methods for estimating linear
  combinations of inverse matrix entries in lattice QCD}},
  \href{https://doi.org/10.1016/0010-4655(94)90004-3}{\emph{Comput. Phys.
  Commun.} {\bfseries 78} (1993) 256}.

\bibitem{Wolff:2003sm}
U.~Wolff, \emph{{Monte Carlo errors with less errors}},
  \href{https://doi.org/10.1016/S0010-4655(03)00467-3,
  10.1016/j.cpc.2006.12.001}{\emph{Comput. Phys. Commun.} {\bfseries 156}
  (2004) 143} [\href{https://arxiv.org/abs/hep-lat/0306017}{{\ttfamily
  hep-lat/0306017}}], [Erratum: Comput. Phys. Commun. 176, 383 (2007)].

\bibitem{Joswig:2022qfe}
F.~Joswig et~al., \emph{{pyerrors: a python framework for error analysis of
  Monte Carlo data}},  \href{https://arxiv.org/abs/2209.14371}{{\ttfamily
  2209.14371}}.

\bibitem{Heitger:2021bmg}
J.~Heitger,  et~al., \emph{{Ratio of flavour non-singlet and singlet scalar
  density renormalisation parameters in $N_\mathrm{f}=3$ QCD with Wilson
  quarks}}, \href{https://doi.org/10.1140/epjc/s10052-021-09387-z}{\emph{Eur.
  Phys. J. C} {\bfseries 81} (2021) 606}
  [\href{https://arxiv.org/abs/2101.10969}{{\ttfamily 2101.10969}}].

\bibitem{Bruno:2022mfy}
M.~Bruno et~al., \emph{{On fits to correlated and auto-correlated data}},
  \href{https://arxiv.org/abs/2209.14188}{{\ttfamily 2209.14188}}.

\bibitem{jureca}
{J\"{u}lich Supercomputing Centre}, \emph{{JURECA: Modular supercomputer at
  J\"{u}lich Supercomputing Centre}},
  \href{https://doi.org/10.17815/jlsrf-4-121-1}{\emph{Journal of large-scale
  research facilities} {\bfseries 4} (2018) }.

\bibitem{hdfcloud}
{J\"{u}lich Supercomputing Centre}, \emph{{HDF Cloud – Helmholtz Data
  Federation Cloud Resources at J\"{u}lich Supercomputing Centre}},
  \href{https://doi.org/10.17815/jlsrf-5-173}{\emph{Journal of large-scale
  research facilities} {\bfseries 5} (2019) }.

\end{thebibliography}\endgroup

\end{document}